\documentclass[aps,prd, tightenlines, footinbib, longbibliography]{revtex4-1}
\usepackage{graphicx}% Include figure files
\usepackage{dcolumn}% Align table columns on decimal point
\usepackage{bm}% bold math
\usepackage[colorlinks,linkcolor=blue,anchorcolor=blue,citecolor=blue,urlcolor=blue]{hyperref}
\usepackage{color}
\usepackage{amsmath}
\usepackage{amsthm}
\usepackage{amssymb}
\usepackage{mathrsfs}
\usepackage{array}
\usepackage{booktabs}
\usepackage{multirow}

\graphicspath{{Figures/}} 

\begin{document}
\baselineskip=0.8 cm
\title{\bf Holographic subregion complexity in unbalanced St\"{u}ckelberg holographic superconductors}

\author{Yu Shi}
\email{shiyu@hhtc.edu.cn}
\author{Chikun Ding}
\author{Yuebing Zhou}
\affiliation{Department of Physics, Huaihua University, Huaihua, Hunan 418008, China}

\author{Qiyuan Pan}
\author{Jiliang Jing}
\email{jljing@hunnu.edu.cn}
\thanks{(Corresponding author)}
\affiliation{Department of Physics, Key Laboratory of Low Dimensional Quantum Structures and Quantum Control of Ministry of Education, and Synergetic Innovation Center for Quantum Effects and Applications, Hunan Normal University, Changsha, Hunan 410081, China}

\begin{abstract}
	\baselineskip=0.6 cm
	\begin{center}
		{\bf Abstract}
	\end{center}

Within the subregion complexity-volume conjecture, we numerically compare holographic subregion complexity (HSC) and holographic entanglement entropy (HEE) for a strip in unbalanced St\"{u}ckelberg holographic superconductors. Varying the St\"{u}ckelberg parameter $\gamma$ yields both second- and first-order transitions. Both observables signal these transitions, but with markedly different robustness. The qualitative HEE signatures persist across strip widths, and the finite part of HEE remains smaller in the superconducting phase than in the normal phase. The HSC is instead strongly width dependent: its temperature trend is opposite to that of HEE at small $\ell$ and agrees with it at large $\ell$. Consequently, the superconducting and normal HSC branches reverse their relative ordering, creating a crossover region where they nearly coincide. There, HSC alone cannot reliably determine the occurrence or order of the transition, and the physical branch must be selected from the grand potential. Thus, HEE provides a more robust diagnostic, whereas HSC is a scale-dependent probe whose interpretation depends explicitly on the subsystem size.
	
\end{abstract}

\maketitle

%\pacs{11.25.Tq, 04.70.Bw, 74.20.-z, 97.60.Lf.}
%\keywords{Holographic subregion complexity; holographic entanglement entropy; unbalanced St\"{u}ckelberg holographic superconductors; phase transition; gauge/gravity duality}

\newpage

\section{Introduction}
The AdS/CFT correspondence, as one of the most prominent realizations of holography, establishes a profound duality between a gravitational theory in a $(d+1)$-dimensional asymptotically anti-de Sitter spacetime and a quantum field theory living on its $d$-dimensional boundary~\cite{AdS-CFT1,AdS-CFT2,AdS-CFT3}. This duality provides a powerful framework for studying strongly coupled systems, since complicated boundary dynamics can be mapped to a weakly coupled gravitational description in the bulk. Over the past two decades, holographic methods have been widely applied to condensed-matter-inspired systems, offering valuable insights into nonperturbative phenomena such as quantum phase transitions, transport properties, and spontaneous symmetry breaking.

Among these applications, holographic superconductors have become a particularly successful framework for modeling superconducting phase transitions at strong coupling \cite{HCM1,HCM2}. In the simplest setup, the normal phase of the boundary theory is dual to a charged AdS black hole, whereas the superconducting phase is associated with the condensation of a charged scalar field in the bulk. Such scalar condensation signals spontaneous symmetry breaking in the boundary theory and leads to a nonvanishing superconducting order parameter below a critical temperature~\cite{HCM3,HCM4}. The holographic superconductor model has been generalized in many directions \cite{HCM5}, among which the St\"{u}ckelberg mechanism provides a flexible way to enrich the phase structure \cite{STU1,STU2}. In particular, by tuning model parameters, one can realize both second-order and first-order phase transitions within the same framework, making this class of models especially suitable for exploring the interplay between strong coupling dynamics and superconducting critical phenomena.

Beyond conventional thermodynamic and transport observables, quantum information measures have also become useful probes of holographic phase transitions. Among them, holographic entanglement entropy, first proposed by Ryu and Takayanagi in Ref.~\cite{Ryu1}, has become an important probe in this context and has been widely used to characterize phase-transition behavior in holographic superconducting systems \cite{HEE1,HEE2,HEE3,HEE4,HEE5,HEE6,HEE7,HEE8,HEE9,HEE10}. Meanwhile, quantum complexity has emerged as another quantity of considerable interest in the study of strongly coupled systems \cite{QC}. Within holography, two representative proposals for its gravitational dual have been put forward, namely the complexity-volume (CV) conjecture \cite{CV1,CV2} and the complexity-action (CA) conjecture \cite{CA1,CA2}. For mixed states or spatial subsystems, holographic subregion complexity (HSC) \cite{HSC}, usually defined by the bulk volume enclosed by a boundary subregion and its corresponding Ryu-Takayanagi (RT) surface, therefore provides a natural quantity to compare with HEE.

This comparison is particularly interesting because HEE and HSC, although associated with the same boundary subsystem, often exhibit qualitatively different behaviors \cite{HSC1,HSC2,HSC3,HSC4,HSC5}. HEE has been widely recognized as an effective probe of superconducting phase transitions, whereas HSC may encode additional information beyond entanglement entropy and thereby reveal a richer structure of the system \cite{THC1,THC2}. Motivated by this possibility, HSC has recently been investigated in a variety of extended holographic superconductor models, including $s$-wave models \cite{SW1,SW2,SW3,SW4,SW5,SW6,SW7}, $p$-wave models \cite{PW1,PW2,PW3}, and $d$-wave models \cite{DW}. These studies suggest that HSC is sensitive to the phase structure and critical behavior of holographic superconductors. However, unlike the relatively robust behavior of HEE, the behavior of HSC is far from universal: its temperature dependence, its response to different orders of phase transitions, and even its effectiveness as a diagnostic can vary significantly from one model to another. A previous study of unbalanced holographic superconductors \cite{SW6} further showed that even for second-order phase transitions, the behavior of HSC depends strongly on the subsystem width, and under certain choices of width the phase-transition information cannot be directly read off from HSC. This naturally raises the question of whether similar ambiguities also arise for other types of phase transitions. To address this question, one needs a model in which different orders of superconducting phase transitions can be realized within the same framework. The unbalanced Stückelberg holographic superconductor provides exactly such a setup.

In this paper, we investigate the holographic subregion complexity (HSC) for a strip subsystem in unbalanced St\"{u}ckelberg holographic superconductors and compare it with the holographic entanglement entropy (HEE). By tuning the St\"{u}ckelberg parameter $\gamma$, the model exhibits both second-order and first-order superconducting phase transitions within a unified framework, with the phase structure and transition order determined from the grand potential. This provides an ideal setup to examine whether the subsystem-size dependence of HSC, previously observed in second-order transitions, persists more generally. We show that the subsystem-width-induced ambiguity of HSC is not confined to second-order transitions but also persists for first-order transitions. Overall, while HEE continues to serve as a relatively robust probe of the phase transition, the HSC signal is much more sensitive to the choice of subsystem size, indicating that its diagnostic power is less universal.

This paper is organized as follows. In Sect.~\ref{section2}, we introduce the unbalanced Stückelberg holographic superconductor model and present the corresponding equations of motion. In Sect.~\ref{section3}, we construct the fully back-reacted bulk solutions. In Sect.~\ref{section4}, we analyze the HEE and HSC and present the main numerical results. Finally, we conclude in Sect.~\ref{section5} with a summary and discussion.

\section{The Holographic Model}\label{section2}
We begin by introducing the unbalanced St\"{u}ckelberg holographic superconductor model in four-dimensional asymptotically AdS spacetime. This model extends the minimal unbalanced holographic superconductor \cite{UN1,UN2} by introducing a St\"{u}ckelberg function, which enriches the phase structure and allows superconducting phase transitions of different orders to be realized within the same framework \cite{STUN}. The bulk action is
\begin{equation}
	S=\frac{1}{2\kappa_4^2}\int d^4x\sqrt{-g}\left(R+\frac{6}{L^2}+\mathcal{L}_{\mathrm{matter}}\right),
	\label{eq2.1}
\end{equation}
with
\begin{equation}
	\mathcal{L}_{\mathrm{matter}}
	=-\frac{1}{4}F^2-\frac{1}{4}Y^2-V(|\psi|)-(\partial\psi)^2-\mathcal{F}(\psi)(\partial p-qA)^2.
	\label{eq2.2}
\end{equation}
Here $F=dA$ and $Y=dB$ are the field strengths of the two $U(1)$ gauge fields $A$ and $B$, respectively. The scalar field $\psi$ is charged under $A$ but neutral under the additional gauge field $B$, so that the imbalance is introduced through the extra gauge sector $B$. In the dual boundary theory, the two gauge fields are associated with two conserved currents, while the gauge field $B$ encodes the chemical potential mismatch related to the $U(1)_B$ ``spin'' symmetry. Since the theory is invariant under the local gauge transformation $A\rightarrow A+\partial\Omega(x)$ and $p\rightarrow p+q\Omega(x)$ \cite{STU1}, we use the gauge freedom to set $p=0$. Throughout this paper, we work in units $L=1$ and $2\kappa_4^2=1$. We choose the St\"{u}ckelberg function to be
\begin{equation}
	\mathcal{F}(\psi)=\psi^2+\gamma\psi^4,
	\label{eq2.3}
\end{equation}
which reduces to the minimal unbalanced holographic superconductor at $\gamma=0$. For definiteness, we take the scalar potential to be $V(\psi)=m^2\psi^2$ with $m^2=-2$ and fix the charge parameter as $q=2$. The chosen mass squared lies above the Breitenlohner-Freedman bound \cite{BF}.

To construct the fully back-reacted background, we adopt the planar black hole ansatz
\begin{equation}
	ds^2=-f(r)e^{-\chi(r)}dt^2+\frac{dr^2}{f(r)}+r^2(dx^2+dy^2),
	\label{eq2.4}
\end{equation}
together with
\begin{equation}
	\psi=\psi(r),\qquad A_a dx^a=\phi(r)\,dt,\qquad B_a dx^a=v(r)\,dt.
	\label{eq2.5}
\end{equation}
Here $f(r)$ and $\chi(r)$ characterize the back-reacted geometry, while $\phi(r)$ and $v(r)$ are the temporal components of the two gauge fields. The Hawking temperature of the black hole is
\begin{equation}
	T_H=\frac{f'(r_H)e^{-\chi(r_H)/2}}{4\pi},
	\label{eq2.6}
\end{equation}
where $r_H$ denotes the radius of the outer horizon, determined by $f(r_H)=0$. According to the holographic dictionary, $T_H$ is identified with the temperature of the dual boundary system. By varying the action, one obtains the independent equations of motion for the metric and matter fields:
\begin{equation}
	\psi''+\psi'\left(\frac{f'}{f}+\frac{2}{r}-\frac{\chi'}{2}\right)-\frac{V'(\psi)}{2f}
	+\frac{e^{\chi}q^2\phi^2}{f^2}\left(\psi+2\gamma\psi^3\right)=0,
	\label{eq2.7}
\end{equation}
\begin{equation}
	\phi''+\phi'\left(\frac{2}{r}+\frac{\chi'}{2}\right)-\frac{2q^2}{f}\left(\psi^2+\gamma\psi^4\right)\phi=0,
	\label{eq2.8}
\end{equation}
\begin{equation}
	\frac{1}{2}\psi'^2+\frac{e^{\chi}(\phi'^2+v'^2)}{4f}+\frac{f'}{fr}
	+\frac{1}{r^2}-\frac{3}{f}+\frac{V(\psi)}{2f}
	+\frac{e^{\chi}q^2\phi^2}{2f^2}\left(\psi^2+\gamma\psi^4\right)=0,
	\label{eq2.9}
\end{equation}
\begin{equation}
	\chi'+r\psi'^2+r\frac{e^{\chi}q^2\phi^2}{f^2}\left(\psi^2+\gamma\psi^4\right)=0,
	\label{eq2.10}
\end{equation}
\begin{equation}
	v''+v'\left(\frac{2}{r}+\frac{\chi'}{2}\right)=0.
	\label{eq2.11}
\end{equation}
Here the prime denotes differentiation with respect to $r$. 

To solve Eqs.~(\ref{eq2.7})--(\ref{eq2.11}) numerically, we impose regular
boundary conditions at both the horizon and the AdS boundary, with the remaining near-horizon coefficients determined by regular Taylor expansions of the equations of motion. Regularity of the gauge fields at the horizon requires
\begin{equation}
	\phi(r_H)=0,\qquad v(r_H)=0.
	\label{eq2.12}
\end{equation}
Near the AdS boundary, $r\to\infty$, the fields admit the asymptotic expansions
\begin{equation}
	\psi(r)=\frac{\psi_1}{r}+\frac{\psi_2}{r^2}+\cdots,
	\label{eq2.13}
\end{equation}
\begin{equation}
	\phi(r)=\mu-\frac{\rho}{r}+\cdots,\qquad
	v(r)=\delta\mu-\frac{\delta\rho}{r}+\cdots,
	\label{eq2.14}
\end{equation}
\begin{equation}
	f(r)=r^2-\frac{\varpi}{2r}+\cdots,\qquad
	\chi(r)\to 0.
	\label{eq2.15}
\end{equation}

According to the holographic dictionary, the leading and subleading terms in the asymptotic expansions of the gauge fields are identified with the chemical potentials and the corresponding charge densities in the dual field theory. In particular, $(\mu,\rho)$ denote the chemical potential and charge density, while $(\delta\mu,\delta\rho)$ correspond to the chemical potential mismatch and the associated density mismatch. The coefficient $\varpi$, related to the black hole mass, is interpreted as the energy density of the dual field theory. For the scalar field, $\psi_1$ and $\psi_2$ are dual to the source and expectation value of the scalar operator, respectively. To realize spontaneous symmetry breaking, we impose the sourceless condition $\psi_1=0$ and identify the condensate as $\langle \mathcal{O}_2\rangle=\sqrt{2}\,\psi_2$.

In the following, we work in the grand-canonical ensemble with fixed chemical potential. The equations of motion admit the scaling symmetry
\begin{equation}
	r\rightarrow \alpha r,\qquad (t,x,y)\rightarrow (t,x,y)/\alpha,\qquad
	\phi\rightarrow \alpha\phi,\qquad v\rightarrow \alpha v,\qquad
	f\rightarrow \alpha^2 f.
	\label{eq2.16}
\end{equation}
To characterize the imbalance independently of this overall scale, we introduce the dimensionless ratio $\beta=\delta\mu/\mu$. We use the radial scaling symmetry to set $r_H=1$ in the raw numerical integration. Physical results are then reported in chemical-potential units, such as $T/\mu$, $\mu\ell$, $s/\mu$, $c/\mu$ and $\Omega/(V_2\mu^3)$. The independent time normalization is used to impose $\chi(\infty)=0$.

\section{Condensation and phase transition}\label{section3}

With the model at hand, we now study the scalar condensation and the associated phase structure in the unbalanced St\"{u}ckelberg holographic superconductor. We begin with the normal phase, which corresponds to a simple analytic solution to Eqs.~(\ref{eq2.7})--(\ref{eq2.11}) and is characterized by a vanishing scalar field, $\psi=0$. In this case, the St\"{u}ckelberg coupling becomes inactive, and the resulting background is independent of the St\"{u}ckelberg parameter $\gamma$.

The system then reduces to a doubly charged Reissner-Nordström-AdS black hole with metric
\begin{equation}
	ds^2=-f(r)\,dt^2+\frac{dr^2}{f(r)}+r^2(dx^2+dy^2),
	\label{eq3.1}
\end{equation}
where
\begin{equation}
	f(r)=r^2\left(1-\frac{r_H^3}{r^3}\right)+\frac{\mu^2 r_H^2}{4r^2}\left(1-\frac{r}{r_H}\right)(1+\beta^2).
	\label{eq3.2}
\end{equation}
The corresponding gauge fields are
\begin{equation}
	\phi(r)=\mu\left(1-\frac{r_H}{r}\right)=\mu-\frac{\rho}{r},
	\label{eq3.3}
\end{equation}
\begin{equation}
	v(r)=\delta\mu\left(1-\frac{r_H}{r}\right)=\delta\mu-\frac{\delta\rho}{r}.
	\label{eq3.4}
\end{equation}
The Hawking temperature of this normal state solution reads
\begin{equation}
	T=\frac{r_H}{16\pi}\left(12-\frac{\mu^2+\beta^2\mu^2}{r_H^2}\right).
	\label{eq3.5}
\end{equation}

\begin{figure}[ht]
	\center{
		\includegraphics[scale=0.7]{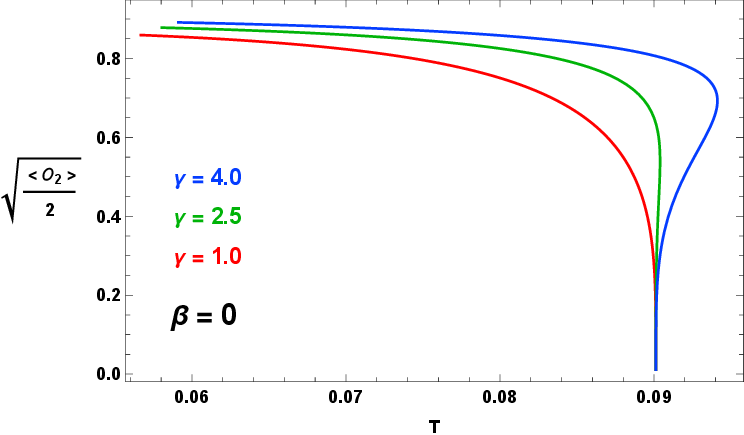}~~
		\includegraphics[scale=0.7]{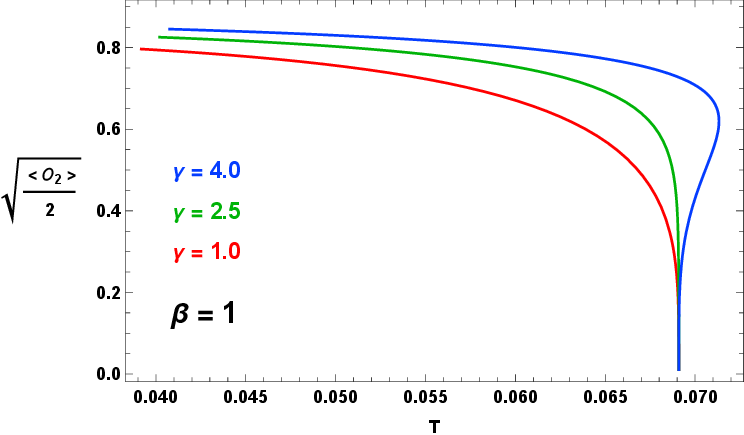}
		\caption{Temperature dependence of the condensate for different values of the St\"{u}ckelberg parameter $\gamma$. The left panel is for the balanced case $\beta=0$, and the right panel is for the unbalanced case $\beta=1$.} \label{fig1} }
\end{figure}

To obtain the superconducting phase with $\psi(r)\neq 0$, we solve the full coupled equations of motion numerically using the shooting method. For numerical convenience, we introduce the dimensionless coordinate $z=r_H/r$. We then study the condensate as a function of temperature and use it as a first probe of the phase structure of the system.

Figure~\ref{fig1} shows the temperature dependence of the condensate for different values of the Stückelberg parameter $\gamma$. Two main features emerge. First, the order of the phase transition is controlled by $\gamma$: as $\gamma$ increases, the system changes from a second-order phase transition to a first-order one. Second, increasing the imbalance parameter $\beta$ lowers the temperature at which the condensate starts to develop, indicating that the imbalance suppresses the onset of condensation.

A particularly interesting case is $\gamma=2.5$. For $\beta=0$, the system undergoes a first-order phase transition, whereas for $\beta=1$, the transition becomes second-order. This indicates that imbalance disfavors first-order behavior, in agreement with the result reported in Ref.~\cite{STUN}.

For a second-order phase transition, the critical temperature $T_c$ can be read off directly from the condensate plot as the temperature at which the condensate starts to develop. From Fig.~\ref{fig1}, we find $T_c\approx 0.090123$ for $\beta=0$ and $T_c\approx 0.069093$ for $\beta=1$. In contrast, for a first-order phase transition the condensate becomes multivalued, and the temperature at which the condensate first appears is not the true critical temperature. Therefore, for first-order phase transitions, the critical temperature cannot be determined directly from the condensate plot and must instead be obtained from the grand potential.

Following the standard procedure of Ref.~\cite{HCM3}, the grand potential in the grand-canonical ensemble is obtained from the renormalized Euclidean on-shell action,
\begin{equation}
	\Omega = T S_E^{\rm on\text{-}shell}.
	\label{eq3.6}
\end{equation}
For the present system, the grand-potential density is
\begin{equation}
	\frac{\Omega}{V_2}=-\frac{\varpi}{2},
	\label{eq3.7}
\end{equation}
where $V_2=\int dx\,dy$ denotes the spatial volume of the boundary system, which is set to unity in the numerical calculation, and $\varpi$ is proportional to the energy density of the dual field theory. For the normal phase, Eq.~(\ref{eq3.6}) can be further reduced to a simple analytic expression,
\begin{equation}
	\frac{\Omega_N}{V_2}
	=
	-r_H^3-\frac{\mu^2}{4}(1+\beta^2)r_H.
	\label{eq3.8}
\end{equation}
Since $\psi=0$ in the normal phase, the St\"{u}ckelberg coupling becomes inactive, and this result is consistent with Ref.~\cite{UN1}.

\begin{figure}[ht]
	\center{
		\includegraphics[scale=0.7]{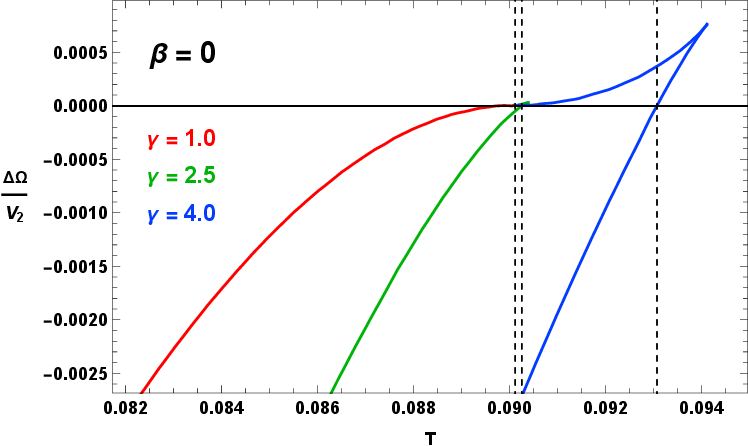}~~
		\includegraphics[scale=0.7]{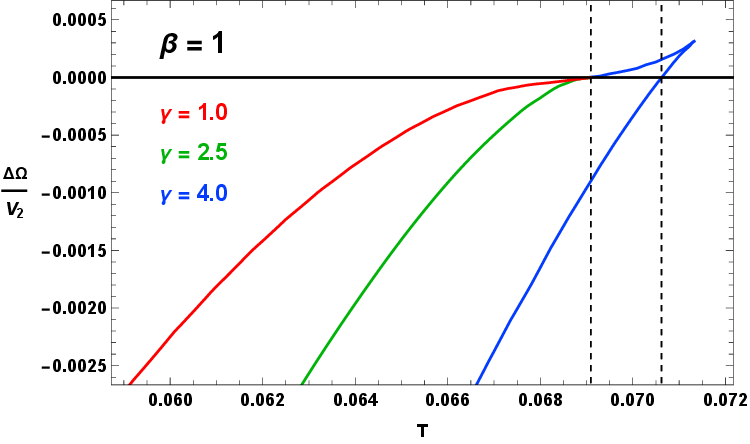}
		\caption{Temperature dependence of the grand-potential difference $\Delta\Omega/V_2$ for different values of the St\"{u}ckelberg parameter $\gamma$. The left panel is for $\beta=0$, and the right panel is for $\beta=1$. The horizontal black line denotes $\Delta\Omega=0$, while the dashed vertical lines indicate the critical temperatures of the phase transitions.} \label{fig2} }
\end{figure}

To determine which phase is thermodynamically preferred at fixed external sources, we therefore consider the grand-potential difference,
\begin{equation}
	\Delta\Omega=\Omega_S-\Omega_N.
	\label{eq3.9}
\end{equation}
Figure~\ref{fig2} shows the temperature dependence of the grand-potential difference. The sign of $\Delta\Omega$ directly determines the thermodynamically favored phase: the superconducting phase is preferred when $\Delta\Omega<0$, whereas the normal phase is favored when $\Delta\Omega>0$. Accordingly, for a first-order phase transition the true critical temperature is determined by the zero crossing $\Delta\Omega=0$, rather than by the temperature at which the condensate first appears.

For $\beta=0$, we find $T_c\approx 0.090259$ for $\gamma=2.5$ and $T_c\approx 0.093077$ for $\gamma=4$. For $\beta=1$, a first-order phase transition remains only for $\gamma=4$, with the corresponding critical temperature $T_c\approx 0.070624$. These results provide a thermodynamic confirmation of the phase structure inferred from Fig.~\ref{fig1}.

More importantly, the disappearance of the zero crossing for $\gamma=2.5$ at $\beta=1$ indicates that the first-order phase transition is no longer realized in this case. For $\gamma=4$, increasing $\beta$ shifts the zero crossing to lower temperature, showing that the imbalance weakens the tendency toward superconducting condensation and lowers the transition temperature. Therefore, imbalance not only suppresses condensation but also makes a first-order phase transition more difficult to realize. Equivalently, a larger St\"{u}ckelberg parameter $\gamma$ is required in the unbalanced system to maintain the first-order nature of the phase transition.

With the phase structure now clarified, we turn in the next section to the holographic entanglement entropy and holographic subregion complexity, and examine how these two nonlocal observables probe the superconducting phase transition.

\section{HEE and HSC of the holographic model}\label{section4}

We first formulate these two nonlocal observables for a straight strip subregion, beginning with the HEE. To this end, we consider on the boundary a straight strip subregion $\mathcal{A}$ with finite width $\ell$ along the $x$-direction and infinite extension along the $y$-direction, namely $-\ell/2\leq x\leq \ell/2$ and $-R/2<y<R/2$, where $R$ is a regulator that will be sent to infinity at the end of the calculation. The corresponding codimension-two extremal surface $\Gamma_{\mathcal{A}}$ is anchored on the boundary of $\mathcal{A}$. In the $z$-coordinate, it starts from the AdS boundary $z=\epsilon$ at $x=\ell/2$, extends into the bulk until reaching the turning point $z=z_\ast$, and then returns symmetrically to the boundary $z=\epsilon$ at $x=-\ell/2$, where $\epsilon$ is the ultraviolet cutoff.

Fixing the time slice and using the bulk metric in Eq.~(\ref{eq2.4}), one obtains the induced metric on the extremal surface $\Gamma_{\mathcal{A}}$ as
\begin{equation}
	ds_{\rm induced}^{2}
	=
	\frac{1}{z^{2}}
	\left\{
	\left[
	1+\frac{1}{z^{2}f(z)}\left(\frac{dz}{dx}\right)^{2}
	\right]dx^{2}
	+dy^{2}
	\right\}.
	\label{eq4.1}
\end{equation}
According to the Ryu--Takayanagi prescription \cite{Ryu1}, the holographic entanglement entropy associated with the subregion $\mathcal{A}$ is given by
\begin{equation}
\mathcal{S}=\frac{{\rm Area}(\Gamma_{\mathcal{A}})}{4G_{N}},
	\label{eq4.2}
\end{equation}
where $G_{N}$ is the Newton constant in the bulk theory. Using the induced metric above, the HEE for the strip geometry can be written as
\begin{equation}
	\mathcal{S}=\frac{R}{4G_{4}}\int_{-\ell/2}^{\ell/2}\frac{dx}{z^{2}}
	\sqrt{1+\frac{1}{z^{2}f}\left(\frac{dz}{dx}\right)^{2}}.
	\label{eq4.3}
\end{equation}

The extremality condition yields
\begin{equation}
	\frac{dz}{dx}
	=
	\frac{1}{z}\sqrt{(z_\ast^{4}-z^{4})f(z)},
	\label{eq4.4}
\end{equation}
where $z=z_\ast$ denotes the turning point of the extremal surface, satisfying $\left.\frac{dz}{dx}\right|_{z=z_\ast}=0$. Integrating this equation, one obtains
\begin{equation}
	x(z)=\int_{z}^{z_\ast}d\tilde z\,
	\frac{\tilde z}{\sqrt{(z_\ast^{4}-\tilde z^{4})f(\tilde z)}},
	\label{eq4.5}
\end{equation}
subject to the boundary condition
\begin{equation}
	x(\epsilon\to 0)=\frac{\ell}{2}.
	\label{eq4.6}
\end{equation}
Substituting the extremality condition back into the entropy functional, the HEE can be rewritten as
\begin{equation}
	\mathcal{S}=\frac{R}{2G_{4}}\int_{\epsilon}^{z_\ast}dz\,
	\frac{z_\ast^{2}}{z^{3}\sqrt{(z_\ast^{4}-z^{4})f(z)}}
	=
	\frac{R}{2G_{4}}\left(s+\frac{1}{\epsilon}\right),
	\label{eq4.7}
\end{equation}
where $s$ denotes the finite part of the holographic entanglement entropy, while the divergent term $1/\epsilon$ represents the usual area-law contribution. In the following, we will focus on the finite part $s$, which contains the physical information relevant to the phase transition.

We now turn to the holographic subregion complexity. According to the subregion CV conjecture \cite{HSC}, the subregion complexity associated with $\mathcal{A}$ is proportional to the bulk volume enclosed by the extremal surface $\Gamma_{\mathcal{A}}$, namely
\begin{equation}
	\mathcal{C}=\frac{V(\Gamma_{\mathcal{A}})}{8\pi L G_{N}}.
	\label{eq4.8}
\end{equation}
Using Eqs.~(\ref{eq4.1}) and (\ref{eq4.8}), the holographic subregion complexity for the strip geometry can be written as
\begin{equation}
	\mathcal{C}
	=
	\frac{R}{4\pi G_{4}}
	\int_{\epsilon}^{z_\ast}\frac{x(z)\,dz}{z^{4}\sqrt{f(z)}}
	=
	\frac{R}{4\pi G_{4}}
	\left(
	c+\frac{F(z_\ast)}{\epsilon^{2}}
	\right),
	\label{eq4.9}
\end{equation}
where $c$ denotes the finite part of the holographic subregion complexity, while $F(z_\ast)/\epsilon^{2}$ represents the divergent contribution. Since the finite part $c$ is independent of the UV cutoff, one may evaluate the complexity at two different cutoffs $\epsilon_1$ and $\epsilon_2$ and then extract $F(z_\ast)$ through
\begin{equation}
	F(z_\ast)
	=
	\frac{4\pi G_4}{R}\,
	\frac{\epsilon_1^2\epsilon_2^2\bigl(\mathcal{C}(\epsilon_1)-\mathcal{C}(\epsilon_2)\bigr)}
	{\epsilon_2^2-\epsilon_1^2}.
	\label{eq4.10}
\end{equation}
In the following, we will focus on the finite part $c$, which contains the physical information relevant to the phase transition.

\begin{figure}[ht]
	\center{
		\includegraphics[scale=0.7]{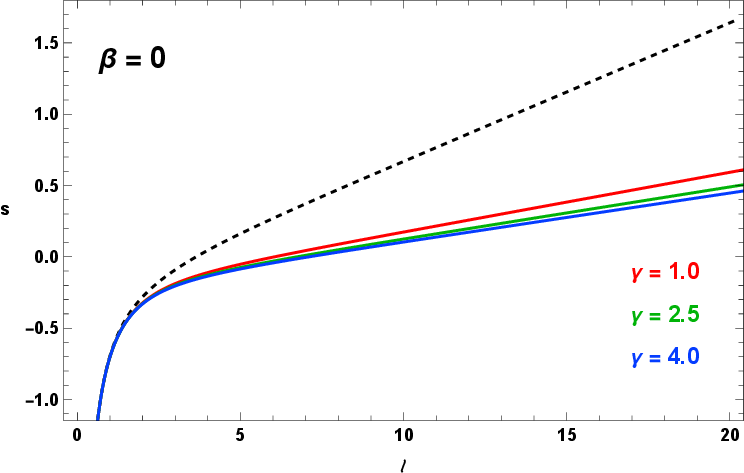}~~
		\includegraphics[scale=0.7]{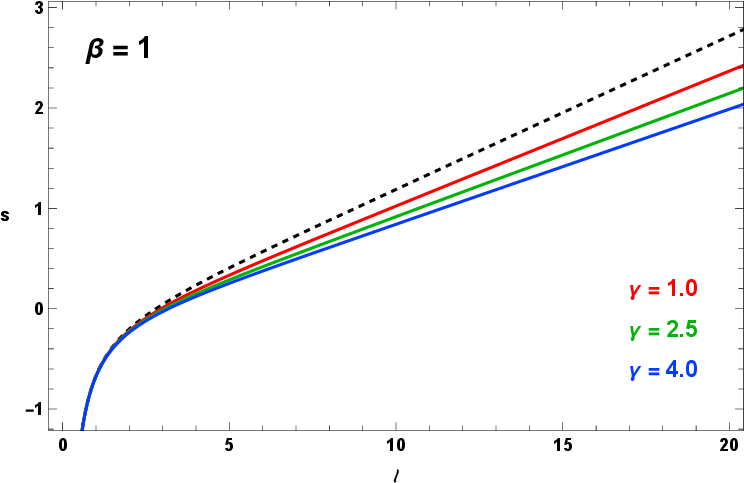}
		\caption{The HEE as a function of the strip width $\ell$ at fixed temperature $T=0.06$ for different values of the St\"{u}ckelberg parameter $\gamma$. The left panel corresponds to $\beta=0$, while the right panel corresponds to $\beta=1$. The dashed black line denotes the corresponding result for the normal phase.} \label{fig3} }
\end{figure}

\begin{figure}[ht]
	\center{
		\includegraphics[scale=0.7]{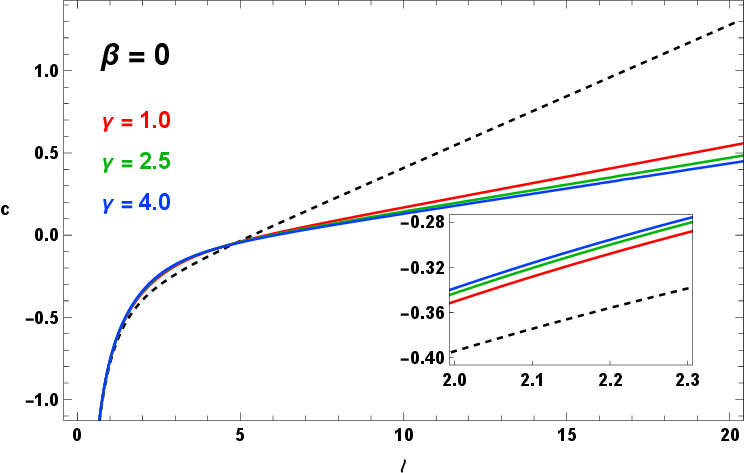}~~
		\includegraphics[scale=0.7]{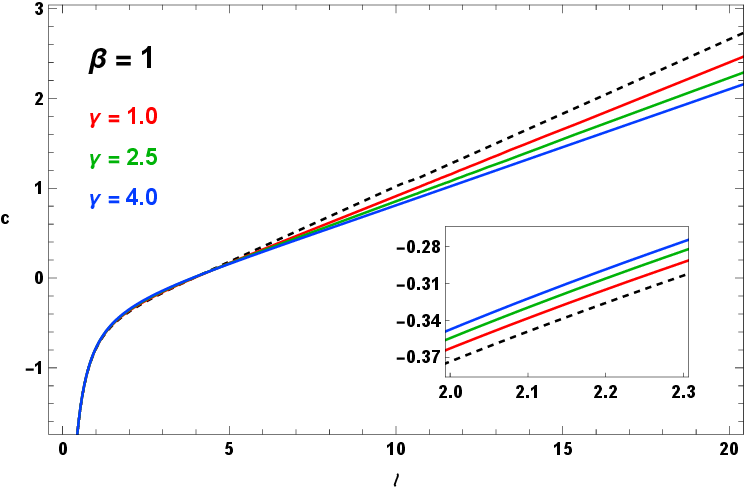}
		\caption{The HSC as a function of the strip width $\ell$ at fixed temperature $T=0.06$ for different values of the St\"{u}ckelberg parameter $\gamma$. The left panel corresponds to $\beta=0$, while the right panel corresponds to $\beta=1$. The dashed black line denotes the corresponding result for the normal phase.} \label{fig4} }
\end{figure}

We first examine the dependence of the HEE and HSC on the strip width at fixed temperature, and then turn to their temperature dependence at representative subsystem sizes. As a representative choice, we fix the temperature at $T=0.06$, which allows a direct comparison between the superconducting and normal phases while avoiding the multivalued region of the superconducting branch. The numerical results for the HEE at this temperature are shown in Fig.~\ref{fig3}. From both panels, one can see that the HEE increases monotonically with the strip width $\ell$. In the large-$\ell$ limit, the finite part of the HEE grows linearly with $\ell$, with the leading contribution determined by the thermal entropy density.

Another important observation is that the HEE in the superconducting phase is always smaller than that in the corresponding normal phase, for both $\beta=0$ and $\beta=1$. This can be understood from the fact that the formation of the condensate reduces the effective degrees of freedom of the system and thus lowers the entanglement entropy. Moreover, at fixed $\beta$, the HEE decreases as the St\"{u}ckelberg parameter $\gamma$ increases. Comparing the two panels, one also finds that the HEE for $\beta=1$ is generally larger than that for $\beta=0$, although the overall qualitative behavior remains unchanged.

\begin{figure}[ht]
\center{
\includegraphics[scale=0.7]{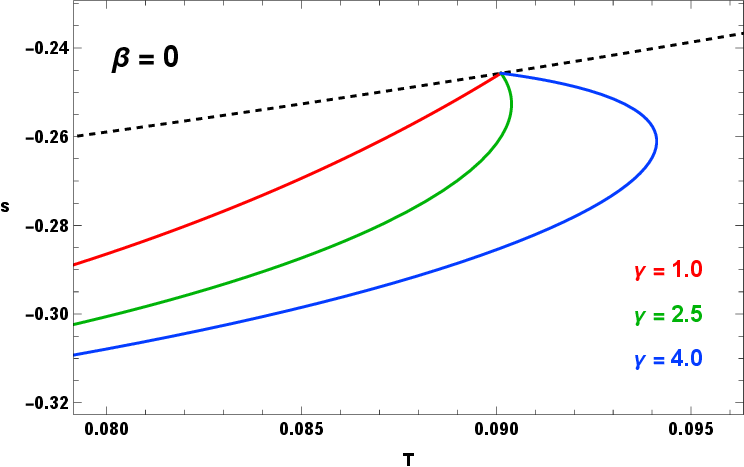}~~
\includegraphics[scale=0.7]{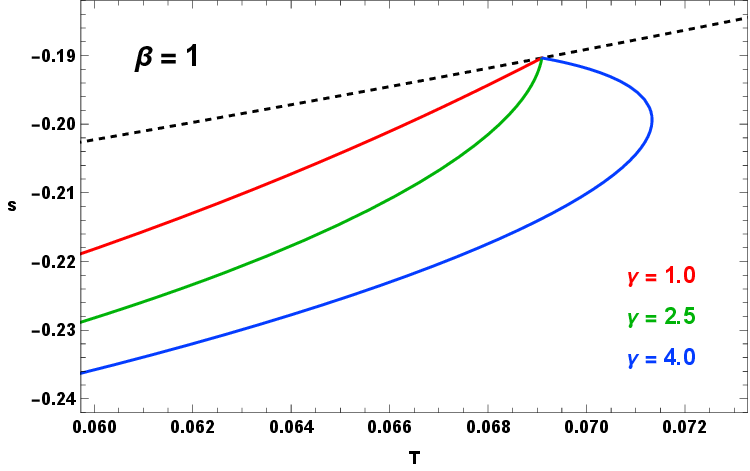}
\caption{Temperature dependence of the HEE at fixed strip width $\ell/2=1$ for different values of the St\"{u}ckelberg parameter $\gamma$. The left panel corresponds to $\beta=0$, while the right panel corresponds to $\beta=1$. The dashed black line denotes the corresponding result for the normal phase.} \label{fig5} }
\end{figure}

\begin{figure}[ht]
	\center{
		\includegraphics[scale=0.7]{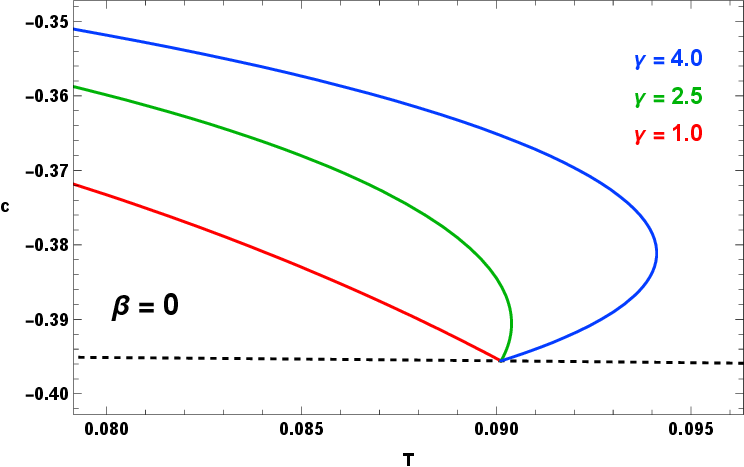}~~
		\includegraphics[scale=0.7]{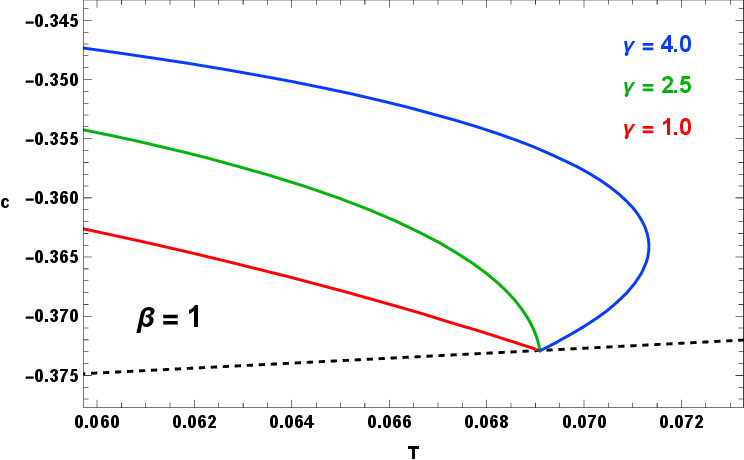}
\caption{Temperature dependence of the HSC at fixed strip width $\ell/2=1$ for different values of the St\"{u}ckelberg parameter $\gamma$. The left panel corresponds to $\beta=0$, while the right panel corresponds to $\beta=1$. The dashed black line denotes the corresponding result for the normal phase.} \label{fig6} }
\end{figure}

\begin{figure}[ht]
	\center{
		\includegraphics[scale=0.7]{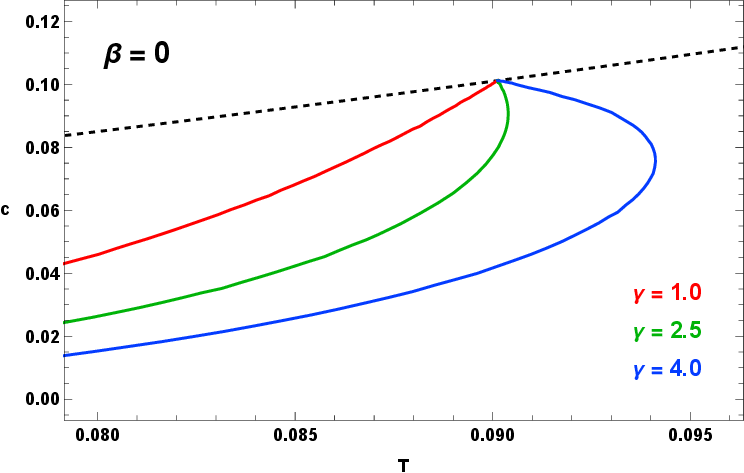}~~
		\includegraphics[scale=0.7]{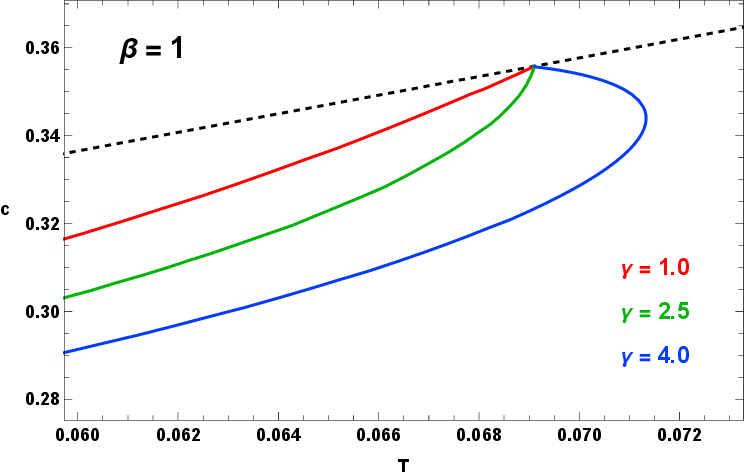}
	\caption{Temperature dependence of the HSC at fixed strip width $\ell/2=12$ for different values of the St\"{u}ckelberg parameter $\gamma$. The left panel corresponds to $\beta=0$, while the right panel corresponds to $\beta=1$. The dashed black line denotes the corresponding result for the normal phase.}\label{fig7} }
\end{figure}

The numerical results for the HSC at the fixed temperature $T=0.06$ are shown in Fig.~\ref{fig4}. Similar to the HEE, the HSC increases monotonically with the strip width $\ell$, and the curves become approximately linear in the large-$\ell$ region, indicating the expected volume-law behavior in the present model.

Taken together, Figs.~\ref{fig3} and \ref{fig4} show that the finite parts of both the HEE and the HSC grow approximately linearly with the strip width $\ell$ for sufficiently large subsystems. However, their detailed responses to the phase structure are markedly different. For the HEE, the superconducting phase always gives a smaller value than the normal phase, and this ordering remains unchanged as $\ell$ varies. By contrast, the HSC exhibits a crossing between the superconducting and normal branches. In the large-$\ell$ region, the HSC in the superconducting phase is smaller than that in the normal phase, whereas for sufficiently small $\ell$ the ordering is reversed. Moreover, at fixed $\beta$, the HSC decreases as the St\"{u}ckelberg parameter $\gamma$ increases. Comparing the two panels, one also finds that the overall value of the HSC for $\beta=1$ is larger than that for $\beta=0$, although the qualitative behavior remains unchanged. These results indicate that, unlike the HEE, the HSC is much more sensitive to the subsystem size and encodes more subtle information about the phase structure. Motivated by these differences, we next turn to the temperature dependence of the HEE and HSC in order to further examine how these two quantities probe the superconducting phase transition.

Figure~\ref{fig5} shows the temperature dependence of the HEE at fixed strip width $\ell/2=1$. It is clear that the HEE provides a robust probe of both the occurrence and the order of the phase transition. For the second-order phase transition, represented here by $\gamma=1.0$, the HEE changes continuously as the temperature decreases. At the critical temperature, it smoothly separates from the normal-phase branch and enters the superconducting branch. In this case, the HEE in the superconducting phase is always smaller than that in the corresponding normal phase, consistent with the expectation that the formation of the condensate reduces the effective degrees of freedom of the system. For the first-order phase transition, the behavior of the HEE becomes qualitatively different. As shown by the curves with larger $\gamma$, the HEE develops a multivalued structure near the transition point and is no longer monotonic in this region. At the critical temperature, the physical branch exhibits a finite jump from the normal phase to the superconducting phase, which is the characteristic signature of a first-order transition. Therefore, the HEE not only signals the phase transition itself but also distinguishes its order through continuity or discontinuity at the critical point.

Comparing the two panels, one finds that these qualitative features remain unchanged when the imbalance parameter is turned on. In particular, the continuous behavior for second-order transitions and the jump behavior for first-order transitions are both preserved, showing that the HEE provides a robust diagnostic
of the phase structure in the unbalanced St\"{u}ckelberg holographic superconductor.

Figure~\ref{fig6} shows the temperature dependence of the HSC at fixed strip width $\ell/2=1$. At this subsystem size, the HSC still provides a clear signal of both the occurrence and the order of the phase transition. For the second-order phase transition, represented here by $\gamma=1.0$, the HSC changes continuously with temperature and smoothly separates from the normal-phase branch at the critical temperature. For the first-order phase transition, corresponding to larger values of $\gamma$, the HSC develops a multivalued structure near the transition point, and the physical branch exhibits a finite jump at the critical temperature. Thus, the HSC can also distinguish the order of the phase transition through its continuity or discontinuity at the critical point.

A notable feature of Fig.~\ref{fig6} is that, at the present strip width, the HSC in the superconducting phase is always larger than that in the corresponding normal phase. This is in sharp contrast to the HEE but is fully consistent with the small-width behavior already observed in Fig.~\ref{fig4}. Comparing the two panels, one further finds that these qualitative features remain unchanged when the imbalance parameter is turned on. This subsystem size lies within the range commonly considered in previous studies, and the resulting behavior is qualitatively consistent with that reported in Refs.~\cite{SW3,PW3}. We will see below that this behavior is not universal but depends sensitively on the subsystem size.

Figure~\ref{fig7} shows the temperature dependence of the HSC at fixed strip width $\ell/2=12$. At this larger subsystem size, the HSC still provides a clear signal of both the occurrence and the order of the phase transition. For the second-order phase transition, represented here by $\gamma=1.0$, the HSC changes continuously with temperature and smoothly branches off from the normal-phase curve at the critical temperature. For the first-order phase transition, corresponding to larger values of $\gamma$, the HSC develops a multivalued structure near the transition point, and the physical branch exhibits a finite jump at the critical temperature. Thus, the HSC can again distinguish the order of the phase transition through its continuity or discontinuity at the critical point.

More importantly, in contrast to the small-width case shown in Fig.~\ref{fig6}, the behavior of the HSC at the present larger strip width becomes qualitatively similar to that of the HEE: the HSC in the superconducting phase is always smaller than that in the corresponding normal phase. This is consistent with the large-$\ell$ behavior already observed in Fig.~\ref{fig4}. Comparing the two panels, one further finds that these qualitative features remain unchanged when the imbalance parameter is turned on. Taken together with the small-width case shown in Fig.~\ref{fig6}, the present result indicates a clear crossover in the temperature dependence of the HSC as the subsystem size increases. In the large-width regime, the behavior of the HSC becomes qualitatively closer to that of the HEE.

 \begin{figure}[ht]
\center{
\includegraphics[scale=0.7]{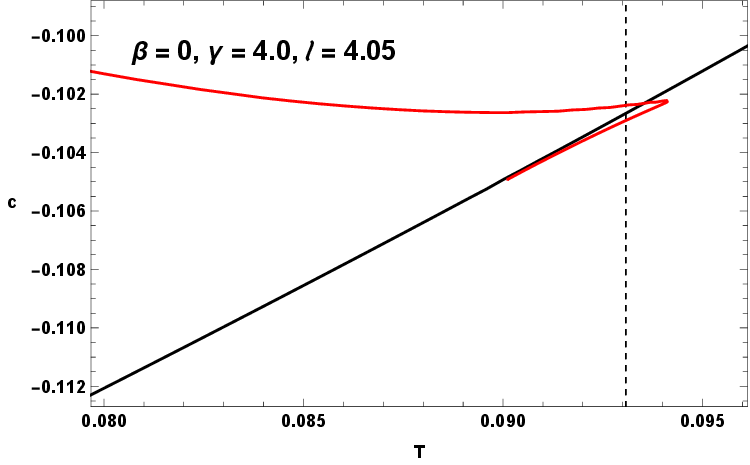}~~
\includegraphics[scale=0.7]{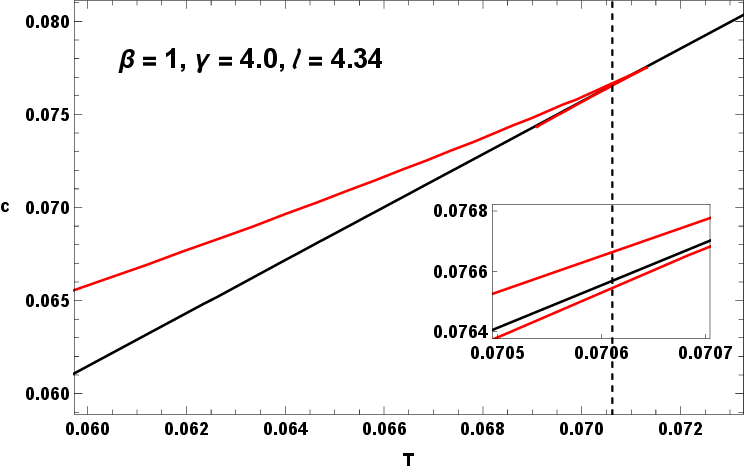}
\caption{Temperature dependence of the HSC for $\gamma=4.0$ at representative strip widths near the crossing region. The left panel corresponds to $\beta=0$ with $\ell=4.05$, while the right panel corresponds to $\beta=1$ with $\ell=4.34$. The red curves denote the superconducting branches, while the black curves denote the corresponding normal-phase branches. The vertical dashed lines mark the corresponding critical temperatures.}
\label{fig8}}
\end{figure}

The numerical results in Fig.~\ref{fig8} further illustrate the strong dependence of the HSC on the subsystem size. To probe the behavior near the crossover between the small- and large-width regimes, we focus on the case $\gamma=4.0$ and choose two representative strip widths close to the crossing region, namely $\ell=4.05$ for $\beta=0$ and $\ell=4.34$ for $\beta=1$. One can see that, in this regime, the behavior of the HSC near the phase transition becomes much more subtle than that shown in Figs.~\ref{fig6} and \ref{fig7}.

A particularly important feature is that, near the critical temperature, the superconducting branch becomes multivalued, and its relative position with respect to the normal branch is no longer definite. As shown in Fig.~\ref{fig8}, one superconducting branch lies slightly above the normal-phase curve, while another lies slightly below it. Therefore, in the vicinity of the crossing region, the HSC no longer provides a direct and unambiguous indication of the phase transition from the relative magnitude of the branches alone. In particular, it becomes difficult to judge, purely from the HSC plot, whether the phase transition has occurred and what its order is.

By contrast, the qualitative HEE signatures remain robust over the same parameter range. The HSC, however, becomes difficult to interpret near the crossing region, where the relevant branches approach and intersect. Consequently, both the occurrence and the order of the phase transition become difficult to discern from the HSC alone. Taken together, these results show that the HEE provides a more robust probe of the phase transition, whereas the diagnostic power of the HSC depends sensitively on the subsystem size and can be substantially weakened near the crossing region.

\section{Summary and discussion}\label{section5}

In this paper, we studied the holographic entanglement entropy and holographic subregion complexity in unbalanced St\"{u}ckelberg holographic superconductors for a straight strip subregion. The St\"{u}ckelberg mechanism allows the system to realize both second-order and first-order superconducting phase transitions for suitable parameter choices. Our results show that the imbalance parameter $\beta$ suppresses scalar condensation, lowers the critical temperature, and weakens the tendency toward a first-order transition.

Our results indicate that the HEE provides a stable characterization of the phase transition. As the temperature varies, it remains continuous for second-order transitions but exhibits a finite jump for first-order ones, thereby clearly distinguishing between the two cases. As the strip width $\ell$ increases, the finite part of the HEE grows approximately linearly in the large-$\ell$ regime and remains below the corresponding normal-state value. These qualitative features persist after the imbalance parameter is introduced, highlighting the robustness of the HEE as a probe of phase transitions in the present model.

By contrast, the HSC depends much more strongly on the subsystem size. Although it also grows approximately linearly with $\ell$ in the large-width regime, its phase-transition signature is not universal. As $\ell$ varies, the relative ordering of the superconducting and normal branches can reverse, indicating that the HSC does not provide a uniform characterization of the transition. For second-order transitions, the expected signature that the HSC remains continuous but develops a discontinuous slope at the critical point can be significantly weakened. In particular, the temperature dependence of the HSC becomes smoother near the critical temperature, and the phase-transition signal becomes less distinct, as already observed in Ref.~\cite{SW6}. Our results further indicate that a similar weakening also occurs for first-order transitions. This demonstrates that, unlike the HEE, the diagnostic power of the HSC is strongly dependent on the subsystem size.

This pronounced $\ell$ dependence merits further discussion. First, the reversal in the relative ordering between the superconducting and normal branches does not appear to be peculiar to the present model, but rather a more generic behavior of the HSC. In the limit $\beta=0$ and $\gamma=0$, our system reduces to the standard $s$-wave holographic superconductor \cite{HCM3}, and the same reversal still occurs. Similar behavior has also been observed in the $p$-wave holographic superconductor model~\cite{PW1} as well as in the $d$-wave holographic superconductor model~\cite{DW}. This suggests that the width-dependent reversal may be a relatively universal feature of holographic subregion complexity in holographic superconductors.

Second, the strip width at which the reversal occurs is not very large in our model. For this reason, the crossover behavior is unlikely to originate from the holographic entanglement plateau effect~\cite{HEP1,HEP2,HEP3}, which is typically associated with the very large-$\ell$ limit, where the RT surface may include the horizon together with the extremal surface of the complementary region. Indeed, along each branch, both the HEE and HSC still exhibit a well-behaved approximately linear dependence on $\ell$, indicating that the observed reversal is more plausibly related to the intrinsic sensitivity of the HSC to geometric details of the bulk spacetime, especially those in the near-horizon region probed by the extremal surface as the strip becomes wider.

Finally, HSC and the complexity defined by the CV conjecture describe,  in principle, different quantities, since the former is associated with a mixed state while the latter is usually interpreted as a pure-state quantity. Even so, as the strip width becomes very large, the RT surface tends to wrap the horizon, suggesting that the corresponding subregion complexity may approach the pure-state complexity more closely. In our results, this tendency is reflected in the large-$\ell$ regime, where the temperature dependence of the HSC becomes qualitatively similar to that of the pure-state complexity obtained from the CV conjecture~\cite{PHC}. A more systematic investigation of this issue would help clarify the geometric information encoded in holographic subregion complexity.

Overall, the main conclusion of our analysis is that robustness and sensitivity do not coincide for these two nonlocal observables. In unbalanced St\"{u}ckelberg holographic superconductors, the HEE provides a stable and transparent characterization of the phase transition, whereas the HSC is much more sensitive to the subsystem size and therefore encodes phase-transition information in a non-universal way. It would be worthwhile to explore whether this feature persists in other holographic superconductor models and under alternative definitions of holographic complexity.

\begin{acknowledgments}
We are grateful to Dr.~Di Wu, Dr.~Sheng Long, Dr.~Yuanceng Xu and Dr.~Hong Guo for useful discussions. This work was supported by the National Natural Science Foundation of China under Grant Nos.~12305063 and 12375047, and by the Hunan Provincial Natural Science Foundation of China under Grant No.~2026JJ30135.
\end{acknowledgments}

\end{document}